\documentclass[runningheads,oribibl]{llncs}

\usepackage{graphicx}
\usepackage{hyperref}
\usepackage{amsmath}
\usepackage{amssymb}
\usepackage{multirow}
\usepackage{xcolor}
\usepackage{url}
\usepackage{placeins}
\usepackage{subfigure}
\usepackage[numbers,sort,compress]{natbib}
\usepackage[normalem]{ulem} 
\newcommand{\PySCF}{\textsc{PySCF}}
\newcommand{\PyFLOSIC}{\textsc{PyFLOSIC}}
\newcommand{\PyFLOSICdev}{\textsc{PyFLOSIC2}}
\newcommand{\ERKALE}{\textsc{ERKALE}}
\newcommand{\Python}{\textsc{Python}}

\newcommand{\fodMC}{\textsc{fodMC}}

\newcommand{\PyCOM}{\textsc{PyCOM}}

\newcommand{\Libxc}{\textsc{Libxc}}

\let\eqref\undefined
\newcommand*{\eqref}[1]{Eq.~(\ref{eq:#1})}
\newcommand*{\tabref}[1]{Table~\ref{tab:#1}}

\newcommand*{\figref}[1]{Fig.~\ref{fig:#1}}

\newcommand*{\secref}[1]{Section~\ref{sec:#1}}
\newcommand*{\secsref}[2]{Sections~\ref{sec:#1} and~\ref{sec:#2}}

\begin{document}
\title{Effect of molecular and electronic
geometries on the electronic density in FLO-SIC}

\titlerunning{Effect of molecular and electronic
geometries}
%
\author{Simon Liebing\inst{1,2}\orcidID{0000-0003-3618-0886} \and
Kai Trepte\inst{3}\orcidID{0000-0003-2214-2467}\and
Sebastian Schwalbe\inst{2}\orcidID{0000-0002-4561-0158}}
\authorrunning{S. Liebing et al.}
\institute{JINR Dubna, Bogoliubov Laboratory of Theoretical Physics,\\ 141980 Dubna, Russia.  \email{science@liebing.cc} \and Institute of Theoretical Physics, TU Bergakademie Freiberg,\\09599 Freiberg, Germany.\and
Stanford University, SUNCAT Center for Interface Science and Catalysis,\\ Menlo Park, CA 94025, USA.
}
\maketitle              
\begin{abstract}

Recently, Trepte et al. [J. Chem. Phys., vol. 155, 2021] \linebreak pointed out the importance 
of analyzing dipole moments in the Fermi-L{\"o}wdin orbital (FLO) self-interaction 
correction (SIC) for cyclic, planar molecules. In this manuscript, the effect of the 
molecular and electronic geometries on dipole moments and polarizabilities is discussed 
for non-cyclic molecules. Computed values are presented for water, formaldehyde, and nitromethane. 
Continuing the work of Schwalbe et al. [J. Chem. Phys. vol. 153, (2020)], we reconfirm 
that systematic numerical parameter studies are essential to obtain consistent results in density functional theory (DFT) and SIC. 
In agreement with Trepte et al. [J. Chem. Phys., vol. 155, 2021], DFT agrees well with experiment for dipole moments, 
while SIC slightly overestimates them.
A Linnett double-quartet electronic geometry is found to be energetically preferred for nitromethane.

\end{abstract}

\section{Introduction/Motivation} 
\label{sec:intro}

Electronic structure methods have become more important over recent years.~\cite{Becke2014_18A301,verma2020_302}
These methods can be used to verify experimental observations.~\cite{Forster2012_856,Pfaff2012_6761,Seidel2013_601,Trepte2017_10020,Trepte2018_25039}
However, the role of electronic structure methods has changed significantly over the years, as they allow to determine properties which are not easily accessible by experiments~\cite{Trepte2015_17122,Ruhlig2017_3963,Taubert2017_942,Schwalbe2017_48}.
Screening for novel materials utilizing purely theoretical and/or computational frameworks saves time, 
work and money.~\cite{Schwalbe2016_8075,Friedrich2019_1,Mehl2021_083608,Trepte2021_630,Friedrich2021_}
The leading methodology is Kohn-Sham (KS) density functional theory (DFT)~\cite{Kohn1965_A1133}, based on its suitable accuracy and reasonable numerical effort. 
Machine learning (ML) strategies are used to speed-up DFT~\cite{Ellis2021_035120} even more or 
to find novel density functional approximations (DFAs)~\cite{Brown2021_2004,Kirkpatrick2021_1385}.
The accuracy of novel DFAs~\cite{Sun2015_036402,Furness2020_8208,Kirkpatrick2021_1385} is getting closer to chemical accuracy.

Some remaining limitations of DFT can be attributed to the so-called self-interaction error (SIE), 
describing artificial interactions of electrons. 
The Perdew-Zunger self-interaction correction (PZ-SIC)~\cite{Perdew1981_5048} approximately removes the one-electron SIE.
It has a long history of successes and failures~\cite{Perdew2015_1}.
In PZ-SIC, the choice of orbitals is important.
Lehtola et al.~\cite{Lehtola2016_3195} showed that PZ-SIC suffers from the local minima problem.
A recent formulation of PZ-SIC utilizes so-called Fermi-L{\"o}wdin orbitals (FLO-SIC)~\cite{Pederson2014_121103,Pederson2015_153,Yang2017_052505,Schwalbe2019_2843,Schwalbe2020_084104}.
FLO-SIC depends on Fermi-orbital descriptors (FODs)~\cite{Schwalbe2019_2843} to construct the localized orbitals used for PZ-SIC~\cite{Trepte2021_224109}. 
These FODs can be imagined as semi-classical electron positions which form an electronic geometry. 
Recently, Trepte et al.~\cite{Trepte2021_224109} showed that one can guide and classify local minima in PZ-SIC 
with the help of special sets of FODs that reflect chemical bonding theories of Lewis~\cite{Lewis1916_762} and Linnett~\cite{Linnett1960_859,Linnett1961_2643}.
The latter is known as Linnett's double-quartet (LDQ) theory.
While typically one is interested in the \emph{variational} total energy of the system, 
Trepte et al.~\cite{Trepte2021_224109} proposed to additionally monitor the dipole moment to classify PZ-SIC solutions.
The dipole moment is one of the most simple descriptor for the electronic density - the key property in any DFA.

In this work we investigate the influence of molecular and electronic geometries as well as a 
properly chosen parameter space on the quality of density-related properties in DFT and FLO-SIC. 
We show that numerical parameters need to be optimized not only for the total energy but also 
for, e.g., the electric dipole moments and/or polarizabilities.
We discuss the results based on small, illustrative and educationally-valuable molecules.

The manuscript is structured as follows.
In the first two sections we outline the theoretical background 
and the computational details, after which we present the major results. 
In the last section we summarize and conclude our findings.

\section{Theoretical background} 
\label{sec:theory}

\begin{figure}
    \centering
    \includegraphics[width=.65\textwidth]{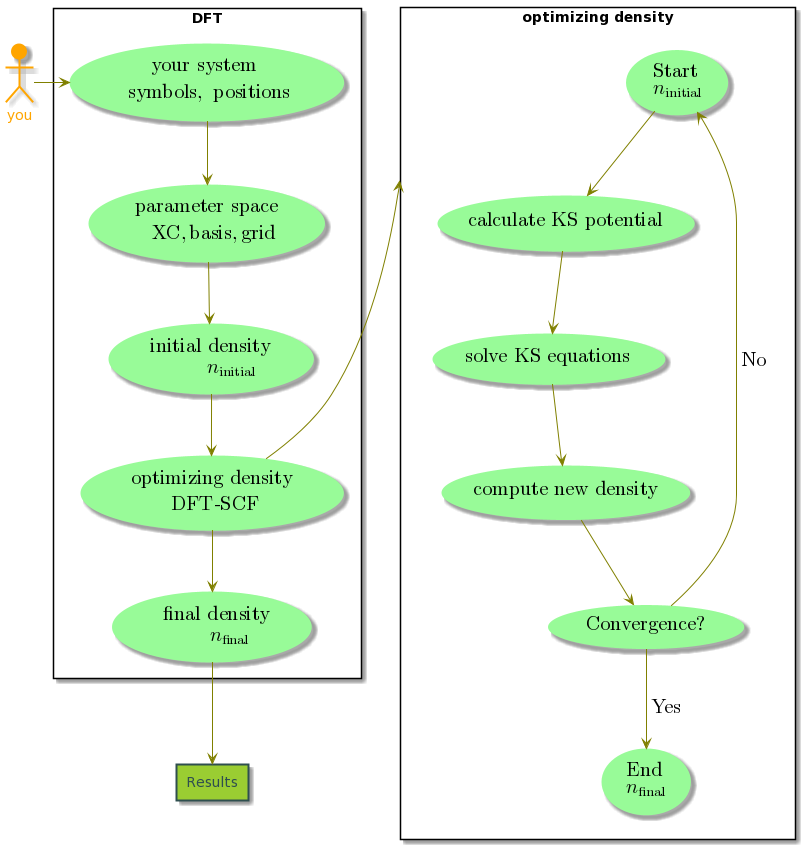}
    \caption{Simplified overview of a DFT calculation.}\label{fig:DFT_SCF_overview}
\end{figure}

KS-DFT, see \figref{DFT_SCF_overview}, is an approximation to solve the Schr{\"o}dinger equation.
The total energy of a system is expressed as a functional of the electron density
\begin{equation}
    E_{\text{KS}}[n^\alpha,n^\beta] = T_{\text{s}}[n^\alpha,n^\beta]+ V[n] + J[n] + K_{\text{XC}}[n^\alpha,n^\beta],\label{eq:EKS}
\end{equation}
where $T_{\text{s}}[n^\alpha,n^\beta]$ is the kinetic energy of the non-interacting system, 
$V[n]$ is the external potential energy, $J[n]$ is the Coulomb functional, $K_{\text{XC}}[n^\alpha,n^\beta]$ is the exchange-correlation (XC) functional,
$n$ is the electron density, and $\alpha$ and $\beta$ indicate spin channels. 

To compute the XC functional
\begin{equation}
    K_{\text{XC}}[n^\alpha,n^\beta] = \int \varepsilon_{\text{XC}}^{\text{hom}}[n^\alpha,n^\beta]n(r) F_{\text{XC}}~\text{d}^{3}r,
    \label{eq:Kxc}
\end{equation}
one needs to evaluate an explicit density integral using a numerical quadrature, see \secref{comp_details}.
Here, $\varepsilon_{\text{XC}}^{\text{hom}}[n^\alpha,n^\beta]$ is the XC energy-density 
of the homogeneous electron gas and $F_{\text{XC}}$ is an XC enhancement factor.

Several approximations exist for the XC enhancement factor, many of which are available in \Libxc{}~\cite{Lehtola2018_1}.
These approximations lead to artificial interactions of electrons with 
themselves; this is called self-interaction (SI). The corresponding SI energy
comes from an incomplete cancellation of the exchange-correlation energy and the Coloumb energy
for one-electron densities $n_{1}^{\sigma}$
\begin{equation}
    E_{\text{SI}}[n_{1}^{\sigma}] = K_{\text{XC}}[n_{1}^{\sigma},0] + J[n_{1}^{\sigma}]. 
\end{equation}

In PZ-SIC, the total $E_{\text{KS}}$ is corrected orbital-by-orbital as 
\begin{equation}
     E_{\text{PZ}} = E_{\text{KS}}[n^{\alpha},n^{\beta}] + E_\text{SIC} =  E_{\text{KS}}[n^{\alpha},n^{\beta}] - \sum_\sigma \sum_{i=1}^{N^\sigma} E_{\text{SI}}[n_{i}^{\sigma}]. \label{eq:EPZ}
\end{equation}

A novel flavor of PZ-SIC is FLO-SIC. This formulation 
utilizes FODs to construct Fermi orbitals (FO).
These FOs are then orthogonalized to become FLOs.
The FODs need to be optimized in the employed numerical parameter space 
using the respective analytical gradients~\cite{Pederson2015_064112}. 
With energy and gradient expressions at hand one can 
study, e.g., ionization potentials, atomization energies 
or barrier heights. 
However, guided by Trepte et al.~\cite{Trepte2021_224109} our focus is not only 
on energies, i.e., \eqref{EKS} and \eqref{EPZ}, but on properties characterizing the density.
Thus, having introduced energy expressions for DFT and PZ-SIC, 
we continue to discuss dipole moments and polarizabilities as 
fingerprints of the electron density. 

Density-related properties can be analyzed using small applied electric fields.
The total energy of a system under an external electrical field $\boldsymbol\varepsilon$ can be written as
\begin{equation}
     E(\boldsymbol\varepsilon) = E_{0} + \sum_{i} \mu_{i} \varepsilon_{i} + \sum_{ij} \alpha_{ij} \varepsilon_{i}\varepsilon_{j} + \mathcal{O}(\boldsymbol\varepsilon^3).
     \label{eq:E_eps}
\end{equation}

Here, $E_{0}$ refers to a ground state energy, e.g, KS-DFT $E_{\text{KS}}$ (see \eqref{EKS}) or PZ-SIC $E_{\text{PZ}}$ (see \eqref{EPZ}). 
From this energy expression we can derive the electric dipole moment as
\begin{equation}
    \mu_{i} = \bigg( \frac{\partial E(\boldsymbol\varepsilon)}{\partial \varepsilon_{i}} \bigg)_{\boldsymbol\varepsilon=0}. 
\end{equation}

Commonly, the dipole moment is directly calculated from the electronic density
\begin{equation}
     \boldsymbol{\mu} = \sum_A Z_A \boldsymbol{R}_A - \int d^3\boldsymbol{r}\, n(\boldsymbol{r})\,\boldsymbol{r},
     \label{eq:mu_ana}
\end{equation}
where $Z_A$, $\boldsymbol{R}_A$, and $n(\boldsymbol{r})$ are nuclear charges and positions and the total electron density, respectively.
Note, we only discuss the electric dipole moment in this work and refer to it simply as \emph{dipole moment}. 
The dipole moment is a measure for the polarity of a system and tells us about the charge separation in this system.
\begin{figure}
    \centering
    \includegraphics[width=.7\textwidth]{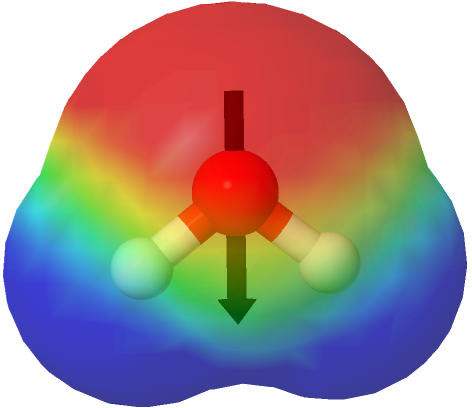}
    \caption{
    Illustration of the dipole moment $\mu$ in the H$_{2}$O molecule using the geometry\,\cite{Hoy1979_1} from CCCBDB\,\cite{Johnson2020_}.
    The molecular electrostatic potential (MEP)~\cite{Scrocco1978_115} was 
    calculated using DFT with LDA-PW, aug-pc-3, and a grid=(200,1454) in \PySCF{}. The visualization was done using \emph{Jmol}~\cite{jmol2022_}.}
    \label{fig:sketch_dm}
\end{figure}
As an example, the molecular electrostatic potential~\cite{Scrocco1978_115} (MEP) and the dipole moment of the H$_{2}$O molecule is visualized in \figref{sketch_dm}.
The dipole points from the more electronegative O atom to the less electronegative H atoms.
This can be directly seen in the coloring scheme of the MEP.

Having the possibility to calculate the total energy of the system under 
the influence of an external electric field allows to study dipole moments using, e.g., a 2-point finite difference (FD) stencil
\begin{equation}
    \mu_{\text{FD},i} = \frac{E(+\varepsilon_{i}) - E(-\varepsilon_{i})}{2\varepsilon_{i}}.
    \label{eq:mu_fd}
\end{equation}

One can also derive the electric polarizability $\alpha_{ij}$ from~\eqref{E_eps} as
\begin{equation}
    \alpha_{ij} = \bigg( \frac{\partial E(\boldsymbol\varepsilon)}{\partial \varepsilon_{i}\varepsilon_{j}} \bigg)_{\boldsymbol\varepsilon=0}. 
\end{equation}
The electric polarizability can be calculated using analytical approaches, e.g., solving the coupled perturbed Hartree-Fock (CPHF) equation~\cite{Handy1984_5031,Sun2020_024109,Smith2020_}.
It describes the tendency of a system to acquire an induced dipole moment in the presence of an external electric field.
Similar to the dipole moment we will refer to the electric polarizability as \emph{polarizability}.

With the dipole moment at hand, i.e., \eqref{mu_ana} 
or \eqref{mu_fd}, one can calculate the directional components of the polarizability tensor as vector components 
\begin{equation}
    [\alpha_{\text{FD},ix},\alpha_{\text{FD},iy},\alpha_{\text{FD},iz}] = \frac{\boldsymbol{\mu}(+\varepsilon_{i}) - \boldsymbol{\mu}(-\varepsilon_{i})}{2\varepsilon_{i}},
    \label{eq:alpha_fd}
\end{equation}
which is a row in \eqref{alpha}. 

To further simplify the characterization of the density, we introduce scalar values. 
The vectorial dipole moment $\boldsymbol{\mu} = (\mu_{x}, \mu_{y}, \mu_{z})$ will be represented as $\mu = |{\boldsymbol{\mu}}|$,
while the tensorial polarizability
\begin{equation}
        \alpha_{ij} = 
          \begin{bmatrix}
            \alpha_{xx} & \alpha_{xy} & \alpha_{xz} \\
            \alpha_{yx} & \alpha_{yy} & \alpha_{yz} \\
            \alpha_{zx} & \alpha_{zy} & \alpha_{zz}  
        \end{bmatrix}
        \label{eq:alpha}
\end{equation}
will be represented as $\alpha = \text{Tr}(\alpha_{ij})/3$. 
As shown by Trepte et al.~\cite{Trepte2021_224109}, dipole moments are sensitive to the chemical bonding
described by the electronic geometry. Moving FODs changes the dipole moment in FLO-SIC. 
Dipole moments in FLO-SIC provide insights into whether the electronic 
density respects the molecular symmetry or not~\cite{Trepte2021_224109}. 

In practical FLO-SIC calculations, the quality of the density is determined 
by the used basis set, numerical quadrature, molecular and electronic geometry. 
Computational details and the used electronic structure codes are discussed next.


\section{Computational details} 
\label{sec:comp_details}

All scripts to produce the data presented in the manuscript are available at \url{https://gitlab.com/opensic/dippo}~\cite{Liebing2022_6246152}.
The calculations were performed with the all-electron 
Gaussian-type orbital (GTO) codes \PySCF{}~\cite{Sun2020_024109} and \PyFLOSICdev{}. 
\PyFLOSICdev{}, see \url{https://gitlab.com/opensic/pyflosic2}, is the successor 
of \linebreak \PyFLOSIC{}~\cite{Schwalbe2020_084104}. It offers a cleaner and more modular code structure 
and can now easily be installed via the \Python{} package manager \emph{pip}.
For calculations of real- and complex-valued SIC, i.e., RSIC and 
CSIC~\cite{Lehtola2014_5324,Lehtola2016_4296,Lehtola2016_3195}, we used the \ERKALE{} code~\cite{Lehtola2012_1572}.
In previous studies~\cite{Schwalbe2020_084104,Trepte2021_224109} we observed 
that the pc-n basis sets~\cite{Jensen2001_9113,Jensen2002_7372} perform well for DFT as well as SIC calculations.
Therefore, for all calculations in this work we use pc-n basis set variants.
All codes use the \Libxc{}~\cite{Lehtola2018_1} library,
offering access to a vast variety of exchange-correlation functionals.
From this library we access LDA-PW~\cite{Perdew1992_13244}, PBEsol~\cite{Perdew2008_136406}, and r$^{2}$SCAN~\cite{Furness2020_8208}.
The used codes are Open-Source codes, meaning 
they are freely available to anyone~\cite{Oliveira2020_024117,Lehtola2021__}.
Open-Source codes enable faster code development, 
re-usable concepts, and versatile tool-boxes.

\PySCF{} and \PyFLOSICdev{} are written in \Python{}, 
where only numerically demanding parts in \PySCF{} are written in C. 
\Python{} is simple and elegant, has a friendly and helpful community, and provides 
various well-maintained libraries. These are only some reasons why it 
is easy for students or non-programmers to start coding with \Python{}.
This allows to solve even non-trivial tasks, like writing a DFT code 
from scratch~\cite{Schulze2021_} in the limited time of a master thesis 
when guided and educated with novel strategies~\cite{Ismail2000_1}.

A numerical quadrature~\cite{Lebedev1999_477}
is needed to evaluate XC properties in DFT and SIC, see \eqref{Kxc} in \secref{theory}.
We will refer to it simply as \emph{grid}.
A typical grid consists of a radial and an angular part. 
Its size is given as a pair of numbers, i.e., the number of radial shells and the number of angular points. 
SIC requires significantly finer grids than DFT~\cite{Vydrov2004_8187,Lehtola2014_5324,Lehtola2016_4296}.
In analogy to \cite{Schwalbe2020_084104,Trepte2021_224109}, we prune the used grids neither for DFT nor for SIC.
This is done as the orbital densities evaluated in SIC are not as smooth as the total density used in DFT~\cite{Shahi2019_174102}, 
thus requiring a finer resolution~\cite{Lehtola2016_4296}.

FLO-SIC has two major variational degrees of freedom, the density matrix (DM) and the FODs.
All FLO-SIC calculations in this work are realized with a two-step 
FLO-SIC SCF cycle, which follows the idea proposed by~\cite{Karanovich2021_014106}.
In FLO-SIC, the initial DM and initial molecular coefficients are typically the 
ones from a DFT calculation. 
The initial FODs can be generated with various procedures, e.g.,
Python-based center of mass \PyCOM{}, Fermi-orbital descriptor Monte-Carlo \fodMC{}, or other so-called FOD generators~\cite{Schwalbe2019_2843}. 
All initial FOD configurations used in this work were generated using the \fodMC{}.

Within a full FLO-SIC calculation, the FODs are fully optimized in an inner FOD loop for a given DM.
All SIC properties are calculated for this DM and the respective optimized FODs.
Based on this, the unified Hamiltonian~\cite{Harrison1983_2079,Pederson1984_1972,Lehtola2013_5365,Schwalbe2020_084104} is updated which then provides the next DM in the outer DM loop.
The initial FODs in the inner FOD loop are optimized using the \emph{SciPy} \mbox{L-BFGS-B}~\cite{Broyden1970_76,Fletcher1970_317,Goldfarb1970_23,Shanno1970_647,Nocedal1980_773,Liu1989_503,Byrd1995_1190,Zhu1997_550} optimization algorithm with a maximum force component threshold of $f_\text{max,tol} = 2 \cdot 10^{-4}$~$E_{\text{h}}$/$a_{0}$.
This two-step procedure is repeated until the FOD forces reach $f_\text{max,tol}$ and the DM is not changing anymore.

The computational methods introduced in this section have been applied to calculate dipole moments and polarizabilities. 
The results for water, formaldehyde, and nitromethane are discussed in the next section.

\section{Results}
\label{sec:results}

\subsection{Sisyphus rock: The importance of grid and basis set size}
\label{sec:sisyphus}

For the correct description of density-related properties, 
it is important to converge the used numerical parameters space
consisting of grid and basis set. 
Systematic parameters studies have been performed utilizing CCCBDB 
molecular geometries~\cite{Johnson2020_}, and FODs generated with the \fodMC{} in the case of FLO-SIC to exemplify this.
The determined trends and optimal values should be transferable to other molecular geometries 
and other FOD arrangements. 

We carried out systematic grid convergence tests varying the number of radial shells with a fixed number of angular points 
and vice versa. Detailed information can be found at \url{https://gitlab.com/opensic/dippo}~\cite{Liebing2022_6246152}. 
The DFT values converge at smaller grids than the 
respective SIC values, see \tabref{grid_basis_conv}. 
For \mbox{LDA-PW}, a value of $N_{\text{rad}} = 200$ for the radial shells
gives converged results for both DFT and SIC. 
The angular dependency for DFT as well as for FLO-SIC
is converged at $N_{\text{ang}} = 590$.
However, we use a value of $N_{\text{ang}} = 1454$ to 
resolve all one-electron and total densities 
accurately.

\begin{table}
    \centering 
    \caption{Optimal numerical parameters for DFT and FLO-SIC regarding the molecular geometries reported in CCCBDB\,\cite{Johnson2020_} for H$_{2}$O\,\cite{Hoy1979_1}, CH$_{2}$O\,\cite{Gurvich1989_} and CH$_{3}$NO$_{2}$\,\cite{Hellwege1976_}.}
    \label{tab:grid_basis_conv}
 \begin{tabular}{c|lll|lll}
\multirow{2}{*}{System}     & \multicolumn{3}{c|}{DFT}                           & \multicolumn{3}{c}{FLO-SIC} \ \\  
                            & $n_{\text{rad}}$ & $n_{\text{ang}}$ & basis set   & $n_{\text{rad}}$ & $n_{\text{ang}}$ & basis set \ \\ \hline
H$_{2}$O                    & 100 & 302 & aug-pc-3                              & 150 & 590 & aug-pc-3  \ \\
CH$_{2}$O                   & 100 & 302 & aug-pc-3                              & 200 & 590 & aug-pc-3 \ \\ 
CH$_{3}$NO$_{2}$            & 150 & 590 & aug-pc-3                              & 200 & 590 & aug-pc-3 \ \\  \end{tabular}  
\end{table}

Note that we investigated the convergence of the density in DFT using the dipole moment as well as the polarizability.
Both density fingerprints deliver the same optimal parameters for the tested molecules. Thus, 
for FLO-SIC we only used the dipole moment to determine optimal parameters.

\begin{figure}[ht!!!!]
    \centering
    \includegraphics[width=1.\textwidth]{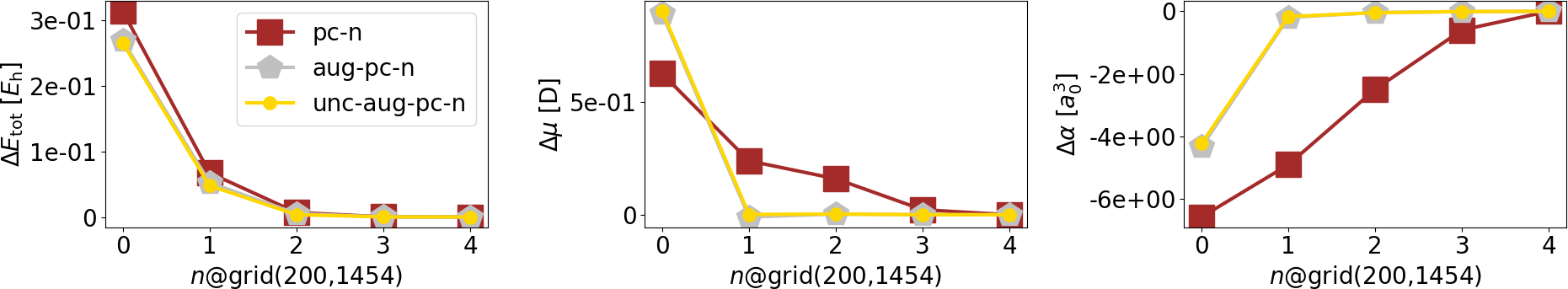}
    \caption{H$_{2}$O (DFT, LDA-PW): Convergence of the total energy $E_{\text{tot}}$, the dipole moment $\mu$ and the polarizability $\alpha$ w.r.t. increasing basis set size for LDA-PW DFT using \PySCF{}. 
    We used pc-n, aug-pc-n, and unc-aug-pc-n with \mbox{n=0-4}~\cite{Jensen2001_9113,Jensen2002_7372}. Each plot shows the difference to the largest used basis set.}\label{fig:water_dft_basis_conv}
\end{figure}

\begin{figure}[ht!!!]
    \centering
    \includegraphics[width=1.\textwidth]{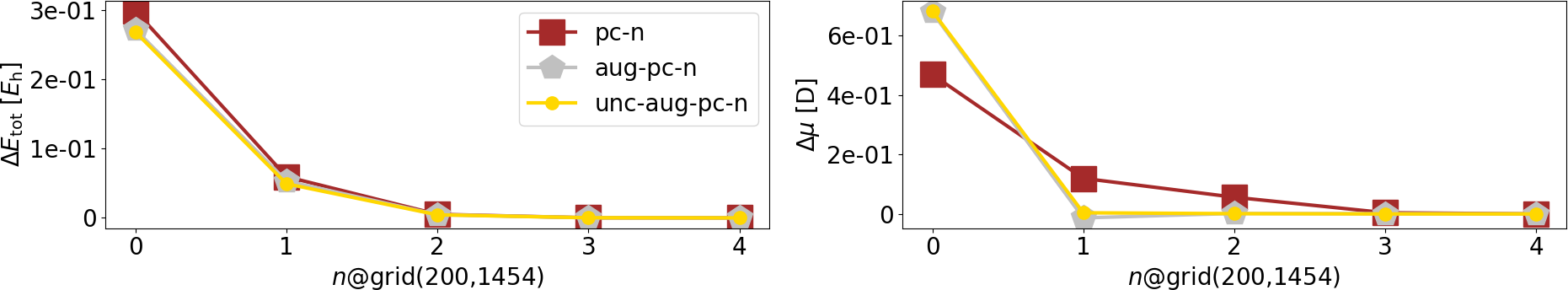}
    \caption{H$_{2}$O (FLO-SIC, LDA-PW): Convergence of the total energy $E_{\text{tot}}$ and the dipole moment $\mu$ w.r.t. increasing basis set size for LDA-PW FLO-SIC using \PyFLOSICdev{}. We used pc-n, aug-pc-n, and unc-aug-pc-n with \mbox{n=0-4}~\cite{Jensen2001_9113,Jensen2002_7372}. Each plot shows the difference to the largest used basis set. Only the density matrix was optimized, while the FODs were fixed.}\label{fig:water_flosic_basis_conv}
\end{figure}

Having an optimal grid of (200,1454) enabled us to determine a suitable basis set. 
The convergence of the basis set for water is 
shown in \figref{water_dft_basis_conv} using DFT and in \figref{water_flosic_basis_conv} using FLO-SIC.
The \mbox{aug-pc-3} basis set shows convergence w.r.t. the basis set size.
Those parameters are optimal for water, formaldehyde, and nitromethane. 
However, such convergence checks need to be done for any molecule - 
a Sisyphus work. 
Otherwise, the meaning of absolute values for dipole moments or polarizabilities are questionable. 

Having established a suitable numerical parameter space, we continue to discuss the influence of molecular geometries on dipole moments and polarizabilities.

\subsection{Pandora's box: The quality of molecular geometries matters}
\label{sec:pandora}

A molecular geometry is needed to perform electronic structure theory calculations;
in case of DFT see \figref{DFT_SCF_overview}. 
While such geometries can be optimized within most theories, 
it is not uncommon to use a fixed molecular geometry to be 
comparable to other approximations or to simply save computational time.

For quantum chemical calculations there exist a vast a variety of 
seemingly promising databases, such as CCCBDB~\cite{Johnson2020_}, 
ChemSpider~\cite{chemspider2022_,Pence2010_1123}, PubChem~\cite{Kim2020_D1388}, and many more.
However, the quality of the geometries in these databases can vary~\cite{Williams2012_685}. 
CCCBDB offers access to a vast variety of molecular geometries. In this work, 
we used the CCCBDB geometries which can be found in the experimental section. 
Note that those geometries are not necessarily experimental ones. 
For example, for water~\cite{Hoy1979_1} the geometry is derived semi-empirically utilizing experimental reference values.
PubChem provides molecular geometries calculated using the MMFF94s~\cite{Halgren1999_720} force field.
For other databases such as ChemSpider it is even not that trivial to find the quality of the 
geometries.

The question of the quality of the molecular geometry 
might be important for other fields as well. For example, 
machine learning~\cite{Fiedler2021_} models might be trained 
on low quality geometries, which could affect the predictability 
of the resulting models.
For SIC calculations, the quality of the molecular geometry is of 
great importance, as the orbital densities are sensitive to the 
underlying molecular geometry.
Molecular geometry optimizations are a standard task for commonly 
used approaches like DFT. However, for more computational 
demanding methods like FLO-SIC, full geometry optimization require 
high computational effort. This is caused by the coupled degrees of freedoms of nuclei and FODs~\cite{Trepte2021_224109}.

The following results are based on the small, educational systems H$_{2}$O, 
CH$_{2}$O, and CH$_{3}$NO$_{2}$. 
We summarized some essential molecular information for those molecules in \tabref{molecular_info}, 
including experimental reference values for dipole moments and polarizabilities.

\begin{table}
    \centering 
    \caption{Molecular information for H$_{2}$O, CH$_{2}$O, and CH$_{3}$NO$_{2}$ with number 
    of atoms, electrons, $\alpha$ and $\beta$ electrons $N_{\text{nuc}}$, $N_{\text{elec}}$, 
    $N_{\alpha}$ and $N_{\beta}$. Experimental references for the dipole moments~\cite{nelson1967} and 
    polarizabilities~\cite{Johnson2020_} are provided.}
    \label{tab:molecular_info}
 \begin{tabular}{cc|llllll}
System & & $N_{\text{nuc}}$ & $N_{\text{elec}}$ & $N_{\alpha}$ & $N_{\beta}$ & $\mu_{\text{REF}}$ [D] & $\alpha_{\text{REF}}$ [$a_{0}^{3}$] \ \\ \hline
water & H$_{2}$O & 3 & 10 & 5 & 5 & 1.85 & 10.13 \ \\
formaldehyde & CH$_{2}$O & 4 & 16 & 8 & 8 & 2.33 & 18.69 \ \\ 
nitromethane  & CH$_{3}$NO$_{2}$ & 7 & 32 & 16 & 16 & 3.46 & 32.39 \ \\  \end{tabular}  
\end{table}

\begin{figure}[ht!!!!]
    \centering
    \includegraphics[scale=0.24]{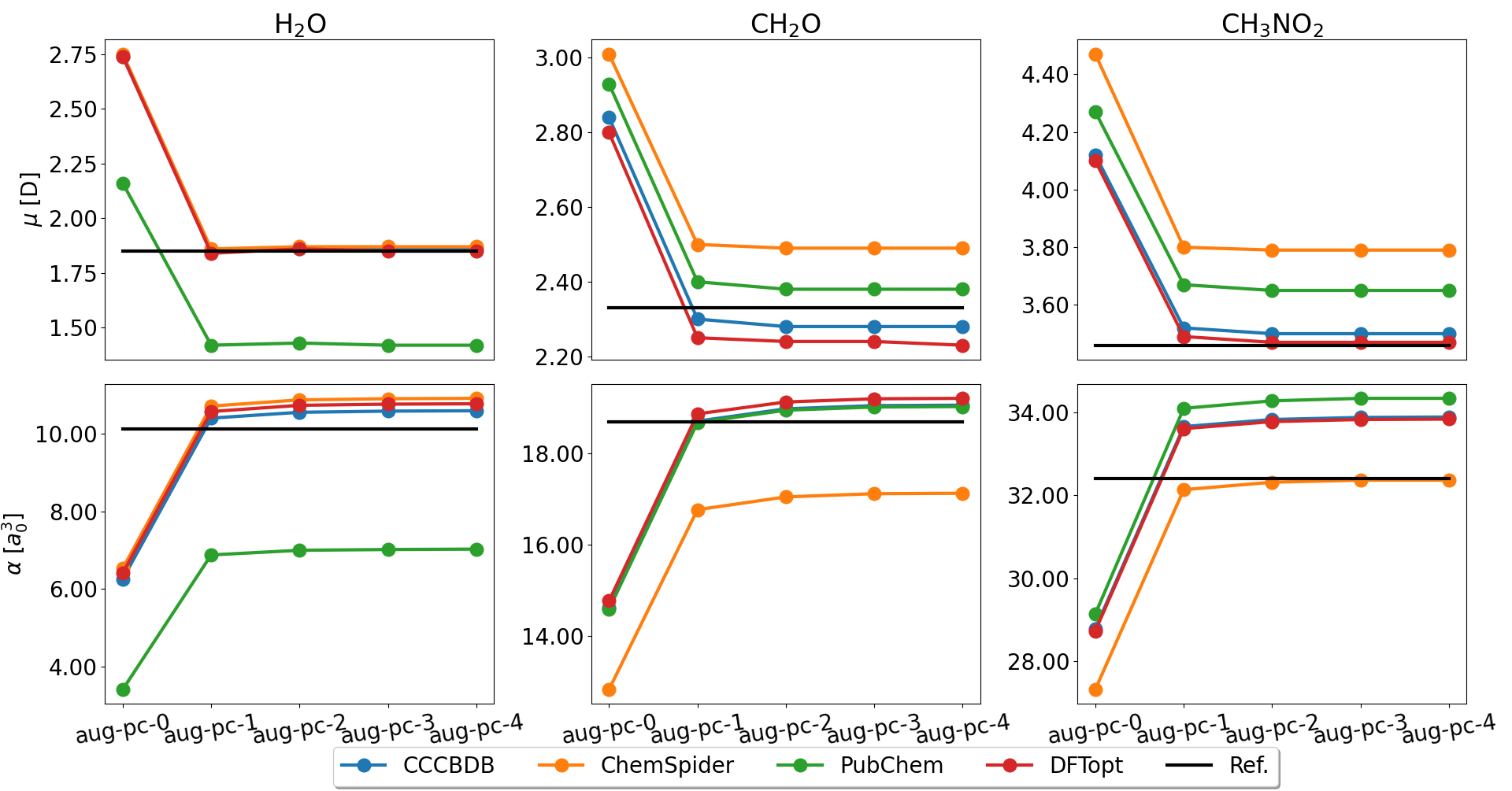}
    \caption{Influence of molecular geometries on density fingerprints, e.g., dipole moments and polarizabilities.
    For these DFT calculations, LDA-PW was used with a grid=(200,1454) in \PySCF{}.
    Molecular geometries are taken 
    from common chemical databases. In addition, DFT optimized geometries were used. 
    These DFT geometry optimizations were started from the CCCBDB geometries utilizing \ERKALE{} with aug-pc-3 and grid=(200,1454).
    The used reference values are provided in \tabref{molecular_info}.}
    \label{fig:nuclei_arrangements}
\end{figure}

The effect of molecular geometries on \emph{density} fingerprints, 
i.e., dipole moments and polarizabilities, is significant even at the DFT level (see \figref{nuclei_arrangements}). 
For the used test systems the molecular geometries from CCCBDB provide the best molecular geometries;
the dipole moments and polarizabilities are close to the values obtain from a DFT optimized geometry. 
The basis set size affects the density-related properties significantly, 
and only the aug-pc-3 basis set provides converged results.
The molecular geometries from PubChem and ChemSpider should be used with care,
as the dipole moments deviate significantly from values obtained from a DFT optimized geometry.

For all calculations, the finite difference approximations, see \eqref{mu_fd} and \eqref{alpha_fd}, agree well with the analytical results. 
The mean error for the dipole moments is 0.00~D, while the respective mean errors 
for the polarizabilities are given in \tabref{delta_approximations}. 
\begin{table}[ht!!!!]
    \centering
    \caption{Finite-difference (FD) errors of polarizabilities in $a_{0}^{3}$ for the chosen step size of $\varepsilon = 10^{-7}$ a.u. utilizing the 2-point FD approximation and LDA-PW DFT.
    The analytical $\alpha$ as calculated with \PySCF{} were used as reference.}
    \label{tab:delta_approximations}
    \begin{tabular}{l|rrrrr}
        Detabase    & aug-pc-0  & aug-pc-1  & aug-pc-2  & aug-pc-3  & aug-pc-4 \ \\ \hline\hline 
        CCCBDB      &    0.00   &    0.29   & $-$0.01   & $-$0.01   & $-$0.01\ \\
        ChemSpider  &    0.00   & $-$0.08   &    0.00   & $-$0.02   & $-$0.00\ \\
        PubChem     & $-$0.01   &    0.02   &    0.02   & $-$0.01   &    0.33\ \\
        DFTopt      &    0.00   &    0.00   & $-$0.03   & $-$0.01   &    0.01\ \\
    \end{tabular}
\end{table}

Accordingly, for the investigated systems and the employed method the chosen value of $10^{-7}$ a.u. for the 
magnitude of $\boldsymbol{\varepsilon}$ regarding the 2-point finite difference approximation delivers reliable numerical 
results. Note that this finding might not be reproducible for other systems using the same value.
Having examined the dependence on the molecular geometry, 
in \secref{damocles} we proceed to investigate electronic degrees of freedom in FLO-SIC, i.e., DM and FODs. 


\subsection{The sword of Damocles: Curse and blessing of approximations} 
\label{sec:damocles}

Approximations are often needed and can be useful to enable the treatment or computation of a specific property at a certain level of theory. 
However, each approximation needs to be carefully investigated regarding the limits of its predictive power.

\subsubsection{Effect of initial FODs}
\label{sec:fods}

\begin{figure}[ht!!!!]
    \centering
    \includegraphics[width=1.1\textwidth]{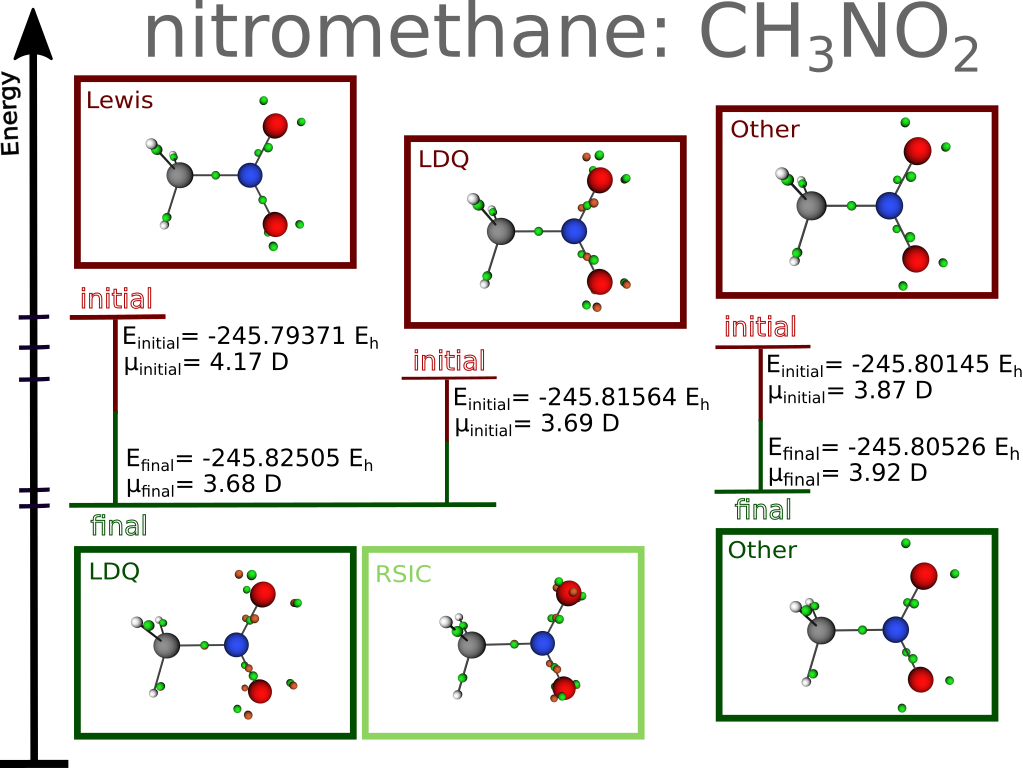}
    \caption{Total energies and dipole moments for various unrestricted FOD configurations for CH$_{3}$NO$_{2}$, evaluated
    at the initial and the final FODs.
    \emph{Other} represents a structure with two N-O double bonds. Any additionally tested structure converged into the LDQ solution.
    Calculations were performed using \PyFLOSICdev{} employing LDA-PW, the aug-pc-2 basis set and a grid=(200,1454).
    In addition, the RSIC center of mass (COM) obtained from \ERKALE{} are shown in the light green box as comparison to the final LDQ FODs.
    Color code: C - gray, H - white, O - red, N - blue, spin-up FOD/COM - green, spin-down FOD/COM - red. There are situations where spin-up and spin-down
    FODs/COMs have the same position. Then, only one color is seen.}
    \label{fig:FODs_nitromethane}
\end{figure}

Continuing the work of Trepte et al.~\cite{Trepte2021_224109}, we show for
CH$_{3}$NO$_{2}$ that it is possible to find several FOD configurations following chemical bonding theories. 
One possible FOD configuration can be based on Lewis theory of bonding. 
There, the FODs form one double N=O and one single N-O bond in the -NO$_{2}$ group, see \figref{FODs_nitromethane}. 
Clearly, there exist two identical Lewis configuration where the double 
and single bonds are exchanged with each other.
These Lewis configuration have N-O bond orders of 2 and 1, respectively.
Regarding LDQ theory, 2 FODs of one spin channel and 1 FOD of the other spin channel are
placed between the N and the O. This leads to bond 
orders of 1.5 in both N-O bonds.
Besides FOD configuration which follow chemical bond theories, other 
FOD configuration are possible. For example, one can generate a configuration with an over-binding N 
atom, placing two N=O double bonds in the molecule.
We denote this FOD configuration as \emph{other}.
Changing the N-O bond order affects the local chemical 
environment, and with that the resulting SIC solution.

In electronic structure theories there exist several possibilities to treat the spin of the system. 
In FLO-SIC, we can do restricted calculations where all electrons are paired, $N_{\alpha} = N_{\beta}$, 
and unrestricted calculations where $N_{\alpha}$ and $N_{\beta}$ can vary. In restricted 
FLO-SIC only one set of FODs is needed, while in unrestricted FLO-SIC two sets of FODs are required. 

Given that the calculations for CH$_{3}$NO$_{2}$ are more computational demanding, we use a grid=(200,1454) and 
the aug-pc-2 basis set. As seen in \secref{sisyphus}, using this basis set
comes with errors in the order of m$E_{\text{h}}$ and mD w.r.t. to the basis set limit.
However, the energy difference of the considered FOD configurations, see \figref{FODs_nitromethane}, are in the order
of $10^{-2}$~$E_{\text{h}}$. The differences in the dipole are in the order of $10^{-1}$~D.
Thus, using the aug-pc-2 basis set should deliver reliable trends.

We performed restricted FLO-SIC calculations for the Lewis configuration of 
CH$_{3}$NO$_{2}$. The FODs converge to an electronic geometry 
which does not follow any bonding theory. 
The double bond FODs are not lying on the N-O bond axis. 
Instead, they are distorted towards the respective O atoms. 
Given this non-symmetric arrangement of FODs, the density of CH$_{3}$NO$_{2}$ becomes non-symmetric. 
This leads to an energy of $E_{\text{final}} = -245.80939$~$E_{\text{h}}$ and a non-symmetric 
dipole moment with an absolute value of $\mu = 4.06$~D.

The effect of various unrestricted FOD configurations for CH$_{3}$NO$_{2}$ is shown in \figref{FODs_nitromethane}. 
The differences in these configurations can already be seen for the initial FODs. 
Only optimizing the density shows significantly different dipole moments, 
and only the LDQ value is close to the experimental value. 
The energy for the initial LDQ arrangement is also the lowest.
Upon full optimization of the density matrix as well as the FODs,
the configuration based on LDQ deliver the lowest energy.
In the unrestricted calculations, both the initial LDQ and the initial Lewis configuration converge 
to a final FOD arrangement which can be characterized via LDQ.
Interestingly, the \emph{other} FODs stay in the their 
configuration, but the final energy is higher than for the LDQ FODs.

LDQ not only gives the lowest energy but also the best SIC dipole moment for CH$_{3}$NO$_{2}$. 
As a highlight, the center of mass (COM) for the optimal RSIC localized orbitals using \ERKALE{} also 
reflect the LDQ chemical bonding motif, see the RSIC box in \figref{FODs_nitromethane}.
In this section, we reconfirm that the choice of the initial FODs significantly influences dipole moments in FLO-SIC.
Next, we describe how the FOD optimization itself can influence dipole moments.

\subsubsection{Effect of FOD optimization}
\begin{figure}
    \centering
    \includegraphics[width=.95\textwidth]{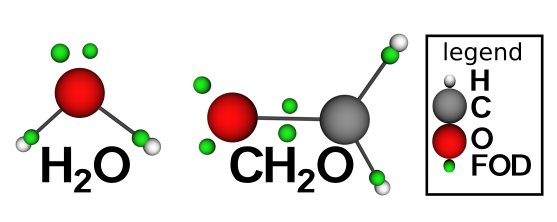}
    \caption{Displaying both the molecular and electronic geometries for H$_{2}$O and CH$_{2}$O. 
    The picture is generated with the \PyFLOSICdev{} graphical user interface (GUI). 
    Note, only H$_{2}$O and CH$_{2}$O are shown as they have only one trivial initial 
    FOD configuration, whereas CH$_{3}$NO$_{2}$ is more complex; see \figref{FODs_nitromethane}.}
    \label{fig:Lewis_FODs}
\end{figure}

We continue the discussion for H$_{2}$O and CH$_{2}$O. 
These two molecules only have one meaningful FOD configuration, see \figref{Lewis_FODs}. 
Having one FOD configuration may allow finding the same local minima 
in a reproducible fashion. 
Thus, those molecules are promising candidates for systematic FOD convergence studies.

\begin{figure}[ht!!!!]
    \centering
    \subfigure[H$_{2}$O]{\includegraphics[scale=0.12]{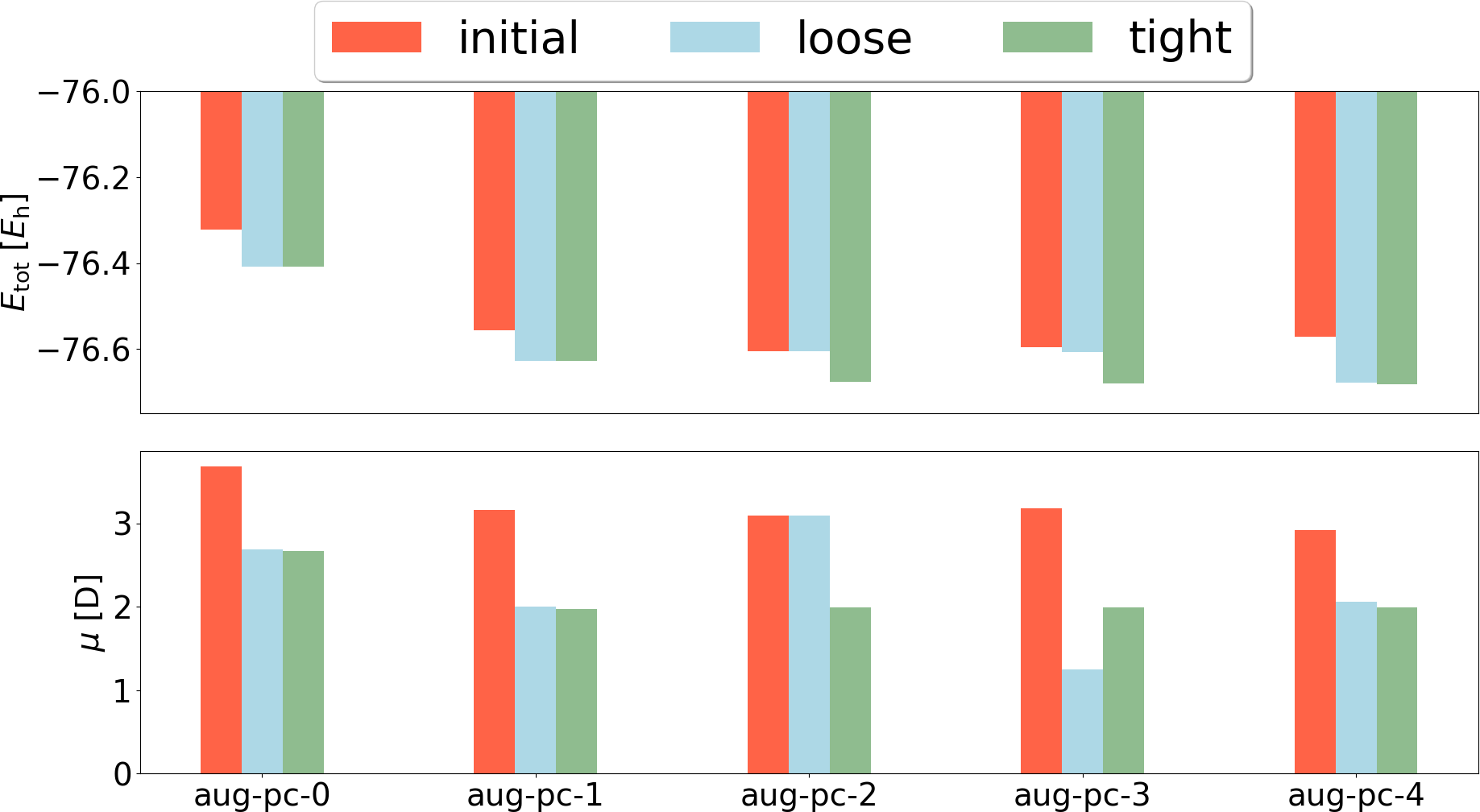}}
    \subfigure[CH$_{2}$O]{\includegraphics[scale=0.12]{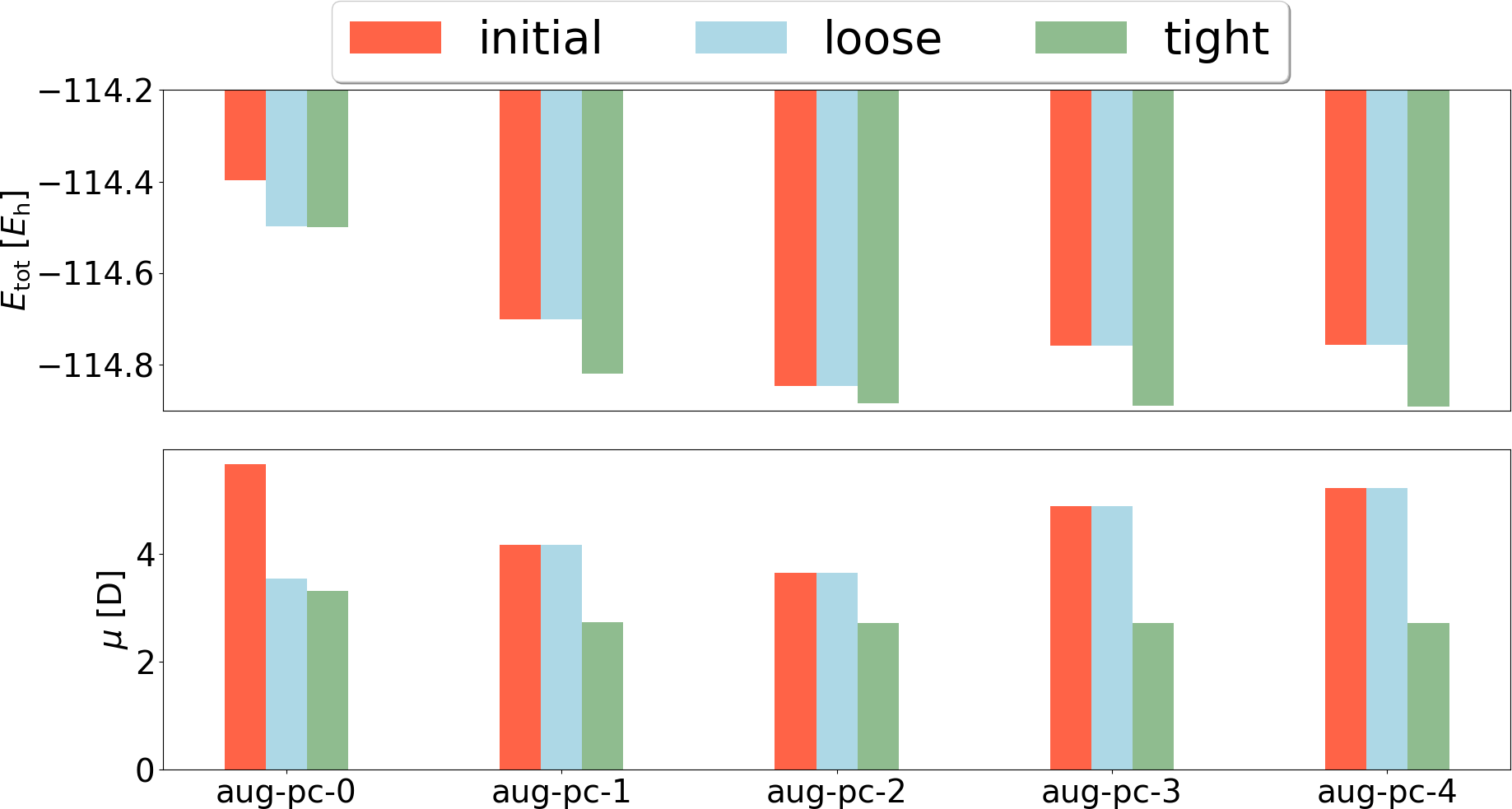}}
    \caption{Effect of FOD optimization vs. basis set size using the LDA-PW XC functional for (a) H$_{2}$O and (b) CH$_{2}$O. 
    The tag \emph{initial} represents FODs generated by the \fodMC{}, \emph{loose} represents FODs 
    with a maximal force criterion of $f_{\text{max,tol}}=5\cdot10^{-3}$~$E_{\text{h}}/a_{0}$,
    and \emph{tight} refers to optimized FODs with $f_{\text{max,tol}}=2\cdot10^{-4}$~$E_{\text{h}}/a_{0}$.
    A grid=(200,1454) was used. Note that when \emph{initial} and \emph{loose} give the same result, the 
    initialized FODs are already at a force threshold of $5\cdot10^{-3}$~$E_{\text{h}}/a_{0}$.}
   \label{fig:tierlevel_vs_flevel}
\end{figure}

In FLO-SIC it is rather easy to make various approximations.
A common approximation is to use fixed FODs which are not 
optimized for the numerical parameter space of the respective calculation. 
Optimizing the FODs is, however, critical to obtain reasonable FLO-SIC solutions. 
Such optimizations are carried out until a specific threshold for the maximum FOD force, 
$f_{\text{max,tol}}$, is reached. 
The influence of the FOD optimization on the total energy and the dipole moment 
for water and formaldehyde is shown in \figref{tierlevel_vs_flevel}.

FLO-SIC values for the total energy and the dipole moment change 
drastically when going from an initial set of FODs to optimized FODs.
Here, optimized FODs are characterized by $f_{\text{max,tol}}=2\cdot10^{-4}$~$E_{\text{h}}/a_{0}$.
Stopping the optimization too early, i.e., at $f_{\text{max,tol}}=5\cdot10^{-3}$~$E_{\text{h}}/a_{0}$
can lead to insufficiently converged densities and energies. This furthermore leads to 
an incorrect prediction of trends; increasing the basis set size should smoothly
converge the dipole moments, see \figref{nuclei_arrangements}. However, with insufficiently optimized FODs these 
trends can be predicted incorrectly, as can be seen in \figref{tierlevel_vs_flevel}.

Accordingly, global statements and generalization of trends are only valid for 
optimized FODs in combination with a sufficient basis set and grid. In the tested cases,
the aug-pc-3 basis set with a grid=(200,1454) deliver converged results, in analogy to \secsref{sisyphus}{pandora}.

As proposed by the authors~\cite{Trepte2021_224109}, monitoring the 
dipole moment is important to classify and analyze SIC solutions.
We showed here that converging the total energy is necessary, but might not be sufficient when one aims to study
density-related properties, i.e., dipole moments or polarizabilities.
All calculations in the previous sections utilized the LDA-PW functional. 
Next, we discuss how changing the exchange-correlation functional influences dipole moments.

\FloatBarrier 

\subsubsection{Effect of exchange-correlation functional}
For H$_{2}$O and CH$_{2}$O, we find that already LDA-PW describes dipole moments 
qualitatively correctly for DFT and SIC. 
However, comparing three exchange-correlation 
functionals the absolute dipole moments differ, see \tabref{comp_mu_XC}.
Note that for CH$_3$NO$_2$ this comparison was not carried,
see \emph{Effect of initial FODs} for more information.


\begin{table}[ht!!!]
    \centering
    \caption{Dipole moments for DFT and FLO-SIC using three XC functionals. 
    Using the aug-pc-3 basis set and a grid=(200,1454) in \PySCF{} for DFT and \PyFLOSICdev{} for FLO-SIC.} 
    \label{tab:comp_mu_XC}
    \begin{tabular}{l|rr|rr}
\multirow{2}{*}{XC functional}  & \multicolumn{2}{c|}{H$_{2}$O}                      & \multicolumn{2}{c}{CH$_{2}$O} \\ 
                                & $\mu_{\text{DFT}}$   & $\mu_{\text{FLO-SIC}}$     & $\mu_{\text{DFT}}$    & $\mu_{\text{FLO-SIC}}$   \ \\ \hline\hline 
        LDA-PW                  & 1.86                 & 1.99                       & 2.28                  & 2.72                  \ \\
        PBEsol                  & 1.82                 & 1.95                       & 2.23                  & 2.65                  \ \\
        r$^{2}$SCAN             & 1.83                 & 1.96                       & 2.31                  & 2.62                   \ \\  
    \end{tabular}
\end{table}


All three pure DFAs agree quite well with the experimental dipole 
reference values, see \tabref{molecular_info}. FLO-SIC tends to 
overshot the dipole moment in order of $10^{-1}$~D w.r.t. the DFT values.
This trend has also been observed in the literature, see~\cite{Trepte2021_224109}.
For DFT, LDA-PW performs best for H$_{2}$O while r$^{2}$SCAN agrees the most for CH$_{2}$O. 
In case of FLO-SIC, the best dipoles are given by PBEsol for H$_{2}$O and by r$^{2}$SCAN for CH$_{2}$O.

In the previous sections we have discussed results for FLO-SIC.
To investigate the influence of a specific flavor of SIC on dipole moments, we compare FLO-SIC, RSIC, and CSIC in the next section.

\subsubsection{Effect of SIC methods}
To verify the used numerical parameter space, i.e., the aug-pc-3 
basis set and a grid=(200,1454),
we carried out RSIC and CSIC calculations using \ERKALE{}.
There is no significant difference between FLO-SIC and RSIC values, see \tabref{comp_mu}. 
This is noteworthy as the results are calculated with two independent electronic structure codes.
Thus, the used numerical parameter space is sufficient to deliver reproducible results.
SIC suffers from the multiple local-minima problem~\cite{Lehtola2016_3195}, see \secref{intro}. Accordingly, we 
recommend to verify FLO-SIC results with independent SIC methods such as RSIC. 
For example for nitromethane, see \emph{Effect of initial FODs}, 
only the LDQ FLO-SIC dipole agrees with the RSIC solution.
FLO-SIC, RSIC and CSIC together deliver a consistent SIC description. 
The advantage of FLO-SIC is the access to bonding information. 
This information allows to easily classify and further analyze PZ-SIC solutions~\cite{Trepte2021_224109}.

\begin{table}[ht!!!]
    \centering
    \caption{Comparison of $\mu$ in D using the aug-pc-3 basis set, a grid=(200,1454) and the LDA-PW functional. Molecular structures from the CCCBDB database\,\cite{Johnson2020_} were used, coming from the following sources: H$_2$O\,\cite{Hoy1979_1}, CH$_{2}$O\,\cite{Gurvich1989_} and CH$_{3}$NO$_{2}$\,\cite{Hellwege1976_}. 
    DFT values are obtained from \PySCF{}, FLO-SIC values are obtained from \PyFLOSICdev{}, and RSIC and CSIC values come from \ERKALE{}.} 
    \label{tab:comp_mu}
    \begin{tabular}{l|rrrr}
        Molecule         & $\mu_{\text{DFT}}$   & $\mu_{\text{FLO-SIC}}$    & $\mu_{\text{RSIC}}$   & $\mu_{\text{CSIC}}$   \ \\ \hline\hline 
        H$_{2}$O         & 1.86                 & 1.99                      & 1.99                  & 2.06                  \ \\
        CH$_{2}$O        & 2.28                 & 2.72                      & 2.72                  & 2.70                  \ \\
        CH$_{3}$NO$_{2}$ & 3.50                 & 3.67                      & 3.70                  & 3.90                  \ \\
    \end{tabular}
\end{table}
In the next section we summarize and conclude our findings.

\section{Summary and Conclusion}

As shown in \secref{sisyphus}, and already stated in earlier works~\cite{Lehtola2014_5324,Schwalbe2020_084104,Trepte2021_224109}, 
SIC needs finer numerical quadrature meshes in comparison to DFT calculations. 
Global statements about the predictive power of SIC are only meaningful using very accurate numerical parameter spaces. 

In this work we show how density-related properties, i.e., dipole moments and 
polarizabilities, can help to determine an appropriate numerical parameter space for DFT and SIC.
For our investigated molecules, the aug-pc-3 basis set and a numerical grid 
of (200,1454) is such an appropriate parameter space. 
Note, even trends can clearly change using other numerical parameters.
While it is mandatory to converge the energy it is not necessarily sufficient 
for the study of density-related properties. 
 
Furthermore, using water, formaldehyde, and nitromethane we show that these density-related 
properties are not only sensitive to the used numerical parameter space. 
They are also significantly influenced by the used molecular geometry, see \secref{pandora}.
Molecular geometries from common chemical database, i.e., CCCBDB, PubChem, or ChemSpider, can deliver very different dipole moments and polarizabilities.
Only molecular geometries from the CCCBDB database deliver reasonable trends in comparison to optimized geometries.

Continuing the work of~\cite{Trepte2021_224109}, for FLO-SIC we showed that density-fingerprints are also sensitive to the 
chemical bonding situation introduced by FODs. We demonstrate that the numerical quality of the FODs, represented by their gradients, 
as well as the choice of initial FODs clearly influences the dipole moments. 
For molecules with non-trivial bonding situations, e.g., nitromethane, it is highly 
recommended to use various initial FOD configurations. This allows to reasonably sample the FOD configuration 
space to determine the most reasonable FOD configuration. Such FOD configurations 
typically follow chemical bonding theories~\cite{Trepte2021_224109}, i.e., Lewis and LDQ.
In case where several FOD configurations are possible, FODs based on LDQ are often superior 
in FLO-SIC. This has been shown in \cite{Trepte2021_224109} and verified here again for nitromethane in \secref{damocles}.

When computational time matters, the simplest density functional approximation, i.e., a local density 
approximation like LDA-PW, provides reasonable trends for dipole moments in FLO-SIC, as seen in \secref{damocles}. 
If computational time is not limited we recommend to use higher rung functional 
like PBEsol or r$^{2}$SCAN to verify and further analyze determined trends. 

We want to emphasis the importance of Open-Science and Open-Source developments, as this work would not be possible without them.


\section{Acknowledgement} 

S. Schwalbe has been funded by the Deutsche Forschungsgemeinschaft (DFG, German Research Foundation) - Project ID 421663657 - KO 1924/9-2.
S. Liebing wants to express his gratitude for Prof. Jens Kortus enabling
the authors to work on this topic.
All authors thank M. Sc. Wanja T. Schulze for his contributions to the \PyFLOSICdev{} code.
We thank both referees, as their comments significantly improved the content and shape of our manuscript.
This work is part of our OpenSIC project, and we thank all members for fruitful discussions. 
We thank the ZIH Dresden for computational time and support. 

%
%

\bibliographystyle{ieeetr}

\bibliography{main}

\end{document}